\documentstyle[prl,aps,twocolumn,epsf]{revtex}

\begin{document}
\draft

\twocolumn[\hsize\textwidth%
\columnwidth\hsize\csname@twocolumnfalse\endcsname

\title{\bf 
On the Kinetic Roughening in Polymer Film Growth 
by Vapor Deposition
}

\author{P. Punyindu and S. Das Sarma}
\address{Department of Physics, University of Maryland, College Park,
MD 20742-4111}

\date{\today}
\maketitle


\vskip 1pc]
\narrowtext

In an interesting recent Letter \cite{one}, Zhao el al. 
report on an experimental investigation of kinetically
rough polymeric (p-xylene) thin film growth using
atomic force microscopy.
The main experimental finding \cite{one}
is a measurement of the effective critical exponents
$\beta$ (growth exponent), $\alpha$ (roughness exponent),
$z$ (dynamical exponent), and $\lambda$ (slope exponent)
characterizing kinetically rough surface growth in
p-xylene.
These exponents are obtained \cite{one}
from a statistical study
of the height-height correlation function of the
evolving growth front as a function of growth time
(or, equivalently, the average film thickness).
The standard dynamic scaling ansatz 
for kinetic surface roughening asserts that the 
root mean square surface width $W(t)$, the surface
correlation length $\xi(t)$, and the average mound
slope \cite{three} $M(t)$ scale as
$W(t) \sim t^\beta$; $\xi(t) \sim t^{1/z}$; 
$M(t) \sim t^\lambda$
in the pre-stationary growth regime characterized
by $t \ll L^z$ (or equivalently $\xi \ll L$)
with the dynamical exponent $z$ given by
$z = \alpha / \beta$.
(An equivalent description follows from the height
correlation function used in Ref. \cite{one}.)
In Ref. \cite{one} the measured exponent values are
reported as $\beta = 0.25 \pm 0.03$; $\alpha = 0.72 \pm 0.05$;
$1/z = 0.31 \pm 0.02$; $\lambda = 0.1-0.3$.
Although the exponents $\alpha$, $\beta$, and $z$
reported in Ref. \cite{one} are completely consistent
with the theoretical predictions \cite{four} of
Lai and Das Sarma (LD) for the conserved nonlinear 
MBE growth universality class, as was noted in Ref. \cite{one},
the authors of Ref. \cite{one} claim that their observations
indicate a novel universality class hitherto unknown
in the literature.
The basis for this claim is the measured non-zero 
value of the slope exponent $\lambda$ ($\approx 0.1-0.3$),
whereas the LD theory \cite{four}, according to Ref. \cite{one},
``predicts stationary growth with a constant local slope''
leading to $\lambda = 0$.

In this Comment we point out that {\it all} the experimentally
measured exponents ($\alpha$, $\beta$, $z$, $\lambda$)
in Ref. \cite{one} are {\it completely consistent with
those of the discrete Das Sarma-Tamborenea (DT) 
limited mobility growth model} \cite{five}
introduced as a minimal model for ideal molecular
beam epitaxy.
In particular, we show in Fig. 1 the measured DT
exponents in 2+1 dimensions using
standard stochastic DT simulations \cite{six,seven}.
It is clear that the DT exponents ($\alpha \approx 0.6$;
$\beta \approx 0.2$; $1/z \approx 0.3$; $\lambda \approx 0.12$)
agree very well with those reported in Ref. \cite{one}.
Our typical simulated DT growth morphology
also agrees remarkably well with the experimental
morphology in Ref. \cite{one}.
Details on the 2+1 dimensional DT model simulations can be 
found in the literature \cite{six,seven}.
The important point we emphasize
\begin{figure}

 \vbox to 8.cm {\vss\hbox to 6cm
 {\hss\
   {\includegraphics{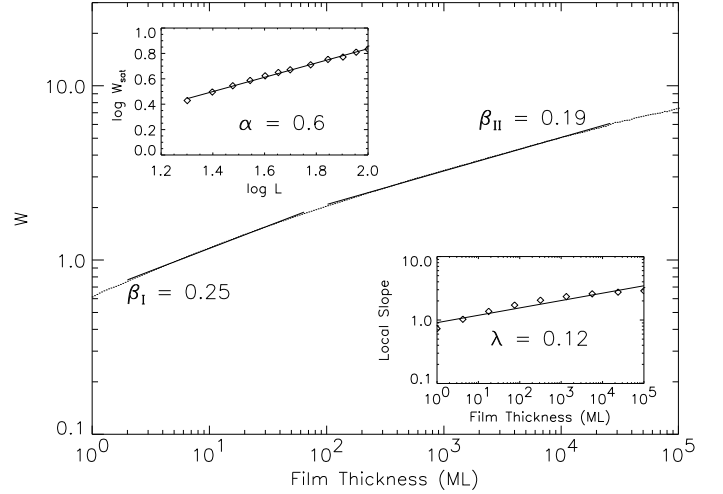}
   }
  \hss}
 }
\caption{
The interface width as a function of deposition time
(average film thickness)
in the $500 \times 500$ DT model.
Top inset: The saturation width $W_{sat}$ vs the substrate size.
Lower inset: The local slope vs time.
}
\label{fig1}
\end{figure}
\noindent
in this Comment is that
DT model produces \cite{seven} a mounded growth
morphology with a finite local slope exponent
($\lambda \neq 0$) 
during a very long-lived pre-asymptotic regime
(even in the absence of any diffusion bias
or Ehrlich-Schwoebel barrier)
with its critical exponents ($\alpha$, $\beta$, $z$)
remaining consistent with LD universality
precisely as observed in Ref. \cite{one}.

We conclude by stating that the fact that the experimental
growth reported in Ref. \cite{one} belongs to the
DT universality class is {\it not} surprising  
because the DT model is a minimal diffusion-driven
vapor deposition growth model.
It is known \cite{eight} that the DT model 
{\it asymptotically} belong to the LD continuum
universality class although there are very interesting
non-asymptotic corrections such as the
non-zero slope exponent \cite{seven}
observed in Ref. \cite{one}.
It will be very interesting to experimentally
observe some of these non-asymptotic corrections
\cite{eight} such as anomalous scaling and
multifractality in the experimental system of Ref. \cite{one}.

This work is supported by the Maryland NSF-MRSEC.

\end{document}